%% file: Paparella.ea.T-ITS_-_Copy.tex
\documentclass[lettersize,journal]{IEEEtran}
\usepackage{amsmath,amsfonts}
\usepackage{algorithmic}
\usepackage{algorithm}
\usepackage{array}
\usepackage[caption=false,font=normalsize,labelfont=sf,textfont=sf]{subfig}
\usepackage{textcomp}
\usepackage{stfloats}
\usepackage{url}
\usepackage{verbatim}
\usepackage{graphicx}
\usepackage{cite}
\usepackage{units}
\usepackage{dsfont}
\usepackage{comment}
\usepackage{booktabs}
\usepackage{marginnote}
\usepackage[dvipsnames]{xcolor}
\input{def.tex}

\usepackage{amsthm}

\newtheorem{theorem}{Theorem}[section]

\newcounter{thm}
\newtheorem{prob}[thm]{Problem}
\def\BibTeX{{\rm B\kern-.05em{\sc i\kern-.025em b}\kern-.08em
		T\kern-.1667em\lower.7ex\hbox{E}\kern-.125emX}}

\newif\ifmargincomments 
\margincommentstrue

\ifmargincomments

\else

\fi

\begin{document}

\title{Electric Autonomous Mobility-on-Demand:\\ Jointly Optimal Vehicle Design and Fleet Operation}

\author{Fabio Paparella, Theo Hofman, Mauro Salazar
\thanks{This publication is part of the project NEON with project number 17628 of the research program Crossover which is (partly) financed by the Dutch Research Council (NWO).}
\thanks{The authors are with the Control System Techonology Section, Eindhoven University of Technology, The Netherlands, \mbox{e-mail:  \tt\footnotesize \{f.paparella,t.hofman,m.r.u.salazar\}@tue.nl}.}}



\maketitle

\begin{abstract}
The advent of autonomous driving and electrification is enabling the deployment of Electric Autonomous Mobility-on-Demand (E-AMoD) systems, whereby electric autonomous vehicles provide on-demand mobility.
Crucially, the design of the individual vehicles and the fleet, and the operation of the system are strongly coupled. Hence, to maximize the system-level performance, they must be optimized in a joint fashion.
To this end, this paper presents a framework to jointly optimize the fleet design in terms of battery capacity and number of vehicles, and the operational strategies of the E-AMoD system, with the aim of maximizing the operator's total profit.
Specifically, we first formulate this joint optimization problem using directed acyclic graphs as a mixed integer linear program, which can be solved using commercial solvers with optimality guarantees.
Second, to solve large instances of the problem, we propose a solution algorithm that solves for randomly sampled sub-problems, providing a more conservative solution of the full problem, and devise a heuristic approach to tackle larger individual sub-problem instances.
Finally, we showcase our framework on a real-world case study in Manhattan, where we demonstrate the interdependence between the number of vehicles, their battery size, and operational and fixed costs.
Our results indicate that to maximize a mobility operator's profit, a fleet of small and light vehicles with battery capacity of \unit[20]{kWh} only can strike the best trade-off in terms of battery degradation, fixed costs and operational efficiency.
\end{abstract}

\input{Sections/Introduction.tex}
\input{Sections/GeneralFramework.tex}

\input{Sections/Sampling.tex}
\input{Sections/Results.tex}
\input{Sections/Conclusions.tex}
\bibliography{main.bib,SML_papers.bib}
\bibliographystyle{IEEEtran}

\begin{IEEEbiography}[{\includegraphics[width=1in,height=1.25in,clip,keepaspectratio]{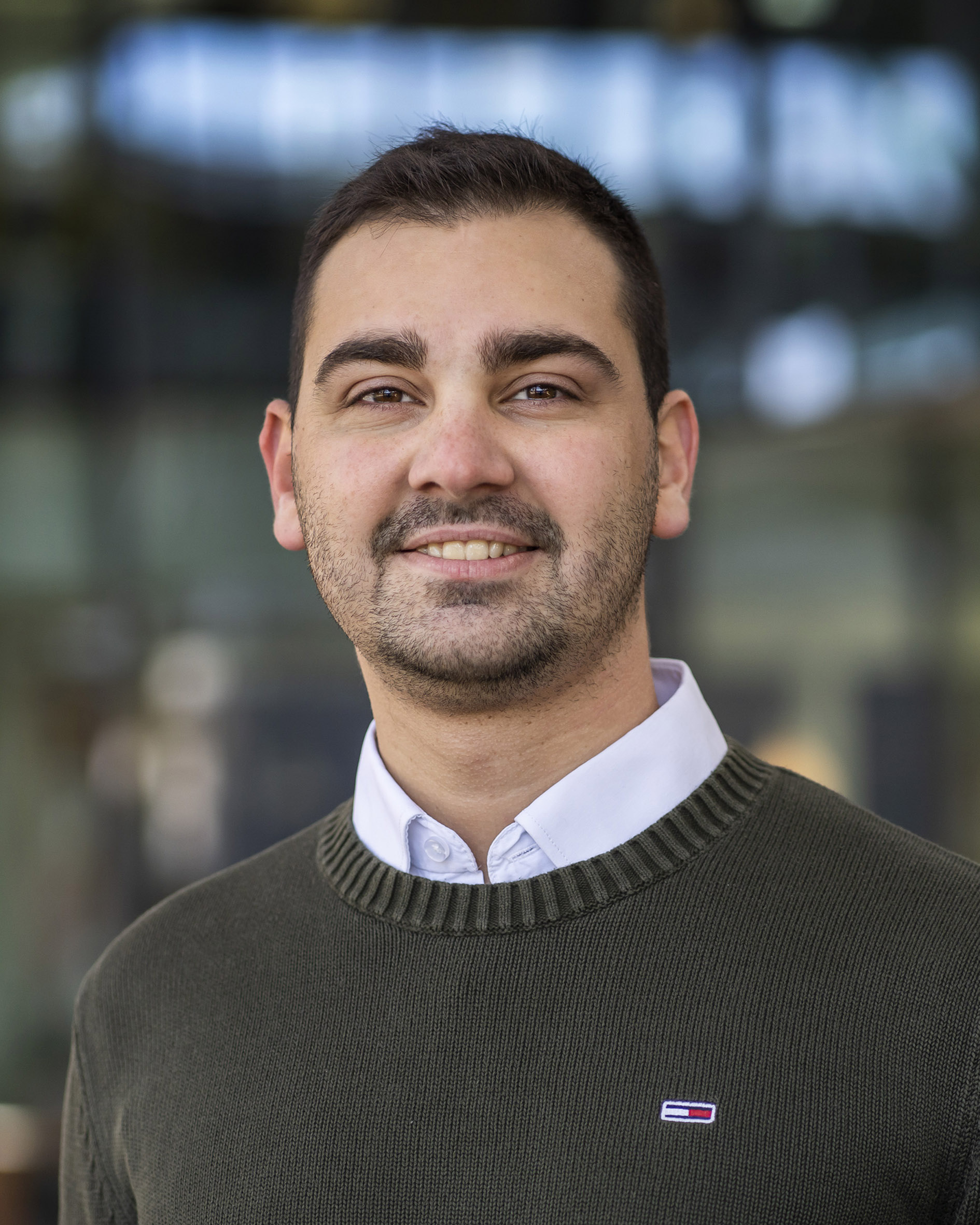}}]{Fabio Paparella} is a Ph.D. student in the Control Systems Technology (CST) section at Eindhoven University of Technology, The Netherlands. He studied mechanical engineering at Politecnico di Milano, Italy, where he received his Bachelor's degree in 2017 and his Master's cum laude in 2020 with a thesis in collaboration with NASA Jet Propulsion Laboratory, California, USA. His research interests include mobility-on-demand, smart mobility, and optimization.
\end{IEEEbiography}

\begin{IEEEbiography}[{\includegraphics[width=1in,height=1.25in,clip,keepaspectratio]{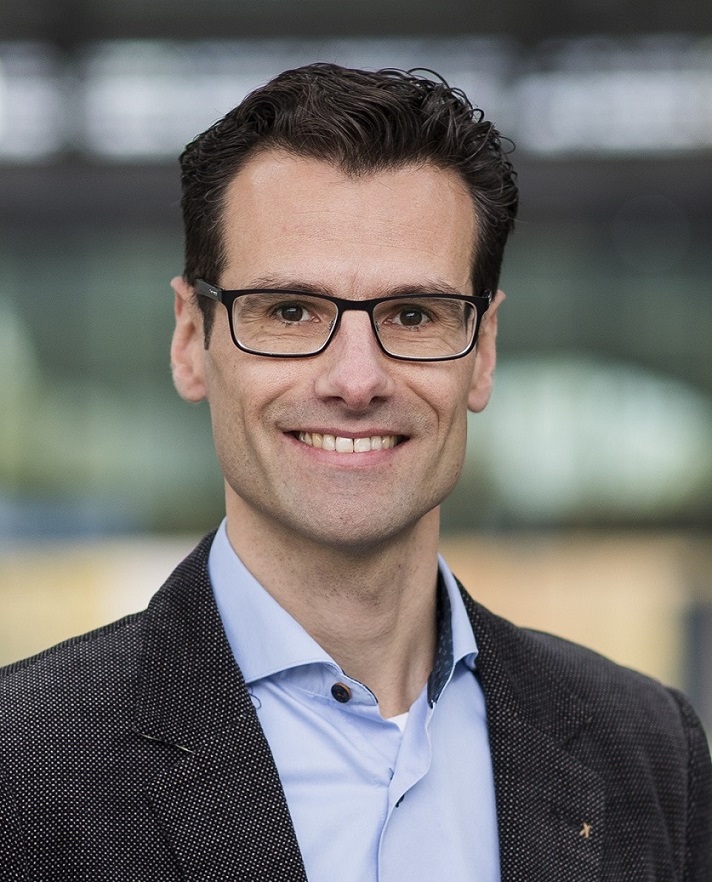}}]{Theo Hofman} was born in Utrecht, The Netherlands, in 1976. He received his M.Sc. (with honors) in 1999 and Ph.D. degree in 2007 both in Mechanical Engineering from Eindhoven University of Technology, Eindhoven. From 1999 to 2003, he was a researcher and project manager with the R\&D Department of Thales Cryogenics B.V., Eindhoven, The Netherlands. From 2003 to 2007, he was a scientific researcher at Drivetrain Innovations B.V., Eindhoven. Since 2010, he is an Associate Professor with the Control Systems Technology group. His research interests are system design optimization methods for complex dynamical engineering systems and discrete topology design using computational design synthesis.
\end{IEEEbiography}

\begin{IEEEbiography}[{\includegraphics[width=1in,height=1.25in,clip,keepaspectratio]{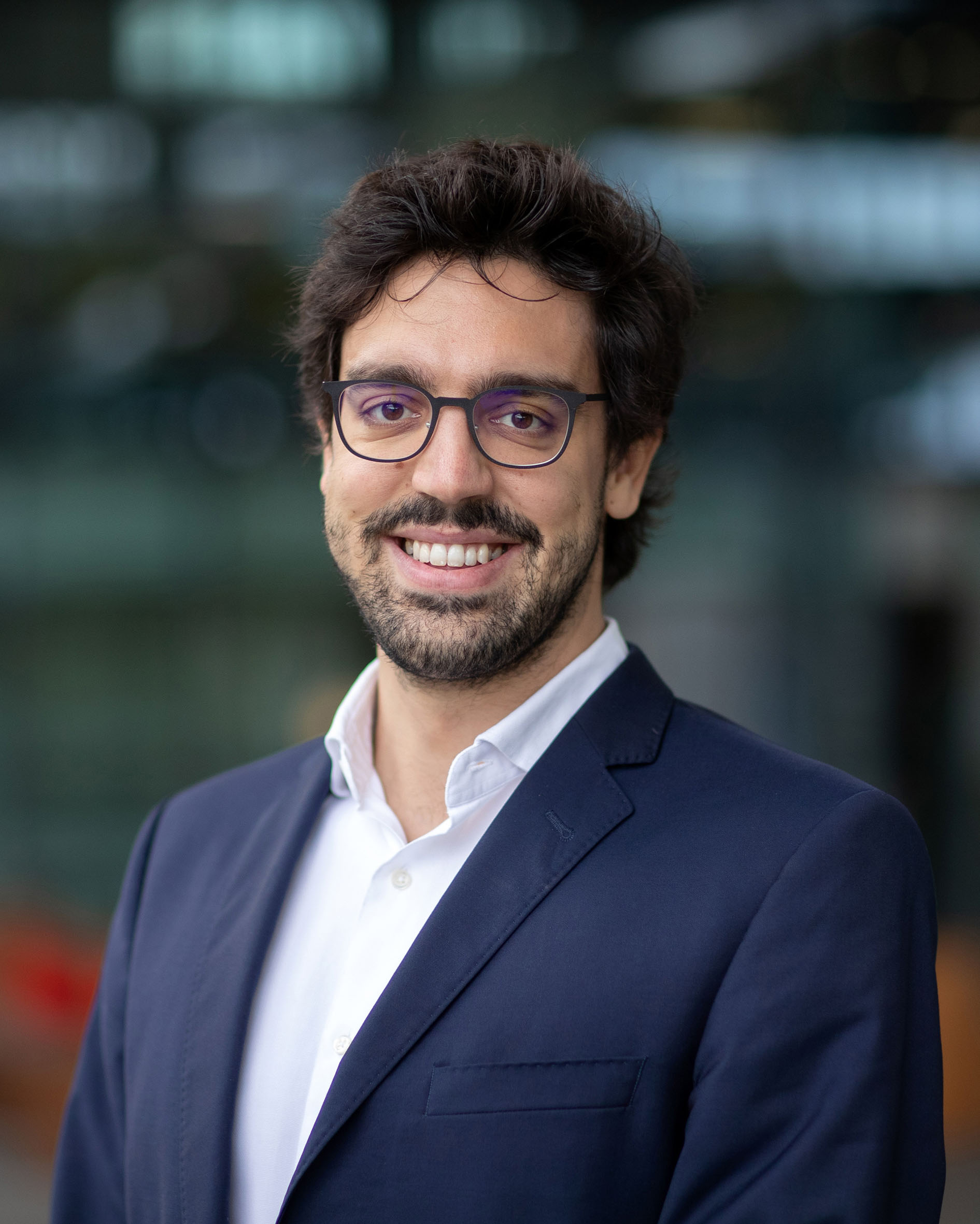}}]{Mauro Salazar} is an Assistant Professor in the Control Systems Technology section at Eindhoven University of Technology (TU/e), and co-affiliated with Eindhoven AI Systems Institute (EAISI). He received the Ph.D. degree in Mechanical Engineering from ETH Zurich in 2019. Before joining TU/e he was a Postdoctoral Scholar in the Autonomous Systems Lab at Stanford University.
Dr. Salazar’s research is focused on optimization models and methods for cyber-socio-technical systems design and control, with a strong focus on sustainable mobility.
Both his Master thesis and PhD thesis were recognized with the ETH Medal, and his papers were granted the Best Student Paper award at the 2018 Intelligent Transportation Systems Conference and at the 2022 European Control Conference.
\end{IEEEbiography}
\vfill

\end{document}

%% file: def.tex
\newcommand{\flexbrac}[1]{\if\relax\detokenize{#1}\relax \else (#1) \fi}
\newcommand{\flexcomma}[1]{\if\relax\detokenize{#1}\relax \else ,#1 \fi}

\newcommand{\abs}[1]{\lvert #1 \rvert}

\newcommand{\cC}{\mathcal{C}}

\newcommand{\cG}{\mathcal{G}}

\newcommand{\cI}{\mathcal{I}}

\newcommand{\cK}{\mathcal{K}}

\newcommand{\cV}{\mathcal{V}}

\newcommand{\sN}{\mathbb{N}}

\newcommand{\sR}{\mathbb{R}}



%% file: Sections/Introduction.tex
\section{Introduction}
\begin{figure}[t]
	\centering
	\includegraphics[width=0.9\linewidth]{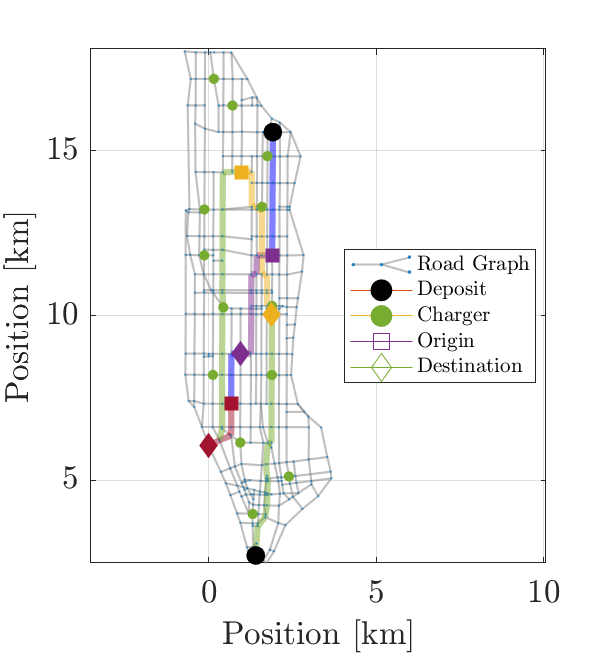}
	\includegraphics[trim={1.5cm 0 0 0},clip,width=1.1\linewidth]{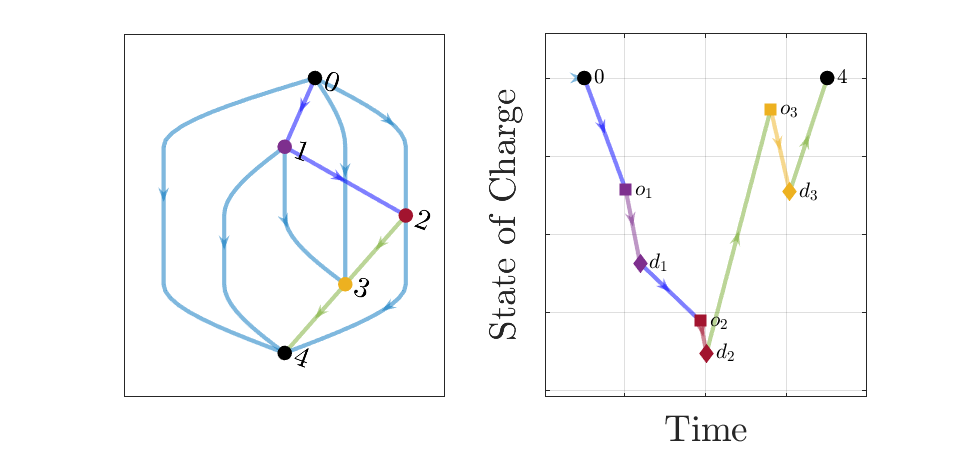}
	\caption{E-AMoD single-vehicle operation on the road graph (top), directed acyclic graph (DAG) representation (bottom-left, with non-selected possibilities in light blue), and energy DAG (E-DAG) representation (bottom-right). Each arc/node on the DAGs represents the correspondingly colored fastest path between the nodes on the road graph. Thereby, green arcs indicate charging during a transition. }
	\label{fig:DAG}
\end{figure}

\IEEEPARstart{M}{obility}-as-a-Service (MaaS) is a solution in the field of mobility that allows users to reserve and pay for several mobility services through a smartphone~\cite{Smith2020} without the need to personally own the used vehicle.
These platforms may address the issues of sustainability and accessibility that mobility systems are currently facing by leveraging opportunities stemming from autonomous driving, connectivity and electrification, for instance with the deployment of Electric Autonomous Mobility-on-Demand (E-AMoD) systems, where electric autonomous vehicles provide mobility services to human users in an on-demand fashion, as shown in Fig.~\ref{fig:DAG}. 
Crucially, the system-level design of an E-AMoD fleet, regarding the number of vehicles, size of their batteries, and charging infrastructure, has a strong influence on both the operational strategies of the fleet and the fixed costs related to its deployment. For instance, a fleet where vehicles are equipped with a large battery size, and therefore have a longer range, provides more flexibility in the charging schedule.
This aspect is important because it enables vehicles to wait until they are near a charging station to minimize the distance driven for recharging purposes, and to assign more vehicles to serving customers during periods of high demand, rather than continuously allocating the charging task to the fleet.
However, larger batteries also increase the cost and weight of each vehicle, leading to higher initial costs, greater energy consumption per distance driven and, given the same charging infrastructure, longer charging times.
In contrast, a smaller battery size may lead to less flexibility in terms of operations at the potential advantage of lower fixed and energy costs.
Therefore, to strike the best trade-off in terms of operational performance and fleet costs, the design and the operation of E-AMoD systems must be jointly studied.
Against this background, this paper proposes a modeling and optimization framework to capture and jointly solve the optimal design and control problem for an E-AMoD system.

\emph{Literature Review:}
This work contributes to the research stream of vehicle-level design jointly optimized with the operation of E-AMoD systems, that we review in the following.

A significant amount of work has been published on the operation of AMoD systems, with a variety of different objectives and methods, as shown by~\cite{ZardiniLanzettiEtAl2022}. Some examples of the latter are queuing-theoretical models~\cite{ZhangPavone2015,ZhangPavone2018,ZhangPavone2014}, agent-based models~\cite{ManleyChengEtAl2014,MartinezViegas2017,FagnantKockelman2014}, vehicle routing problem (VRP) ~\cite{YaoChenEtAl2021b,PavoneFrazzoli2010,PavoneBisnikEtAl2007,TothVigo2014,Laporte1992,PsaraftisWenEtAl2016,PillacGendreauEtAl2013}, and multi-commodity network flow models~\cite{PavoneSmithEtAl2011,SalazarTsaoEtAl2019,PaparellaChauhanEtAl2023,PaparellaSripanhaEtAl2022}.
In particular, network flow models have been leveraged to optimize the operation accounting for a wide range of factors such as congestion-aware routing~\cite{RossiZhangEtAl2017}, intermodality~\cite{SalazarLanzettiEtAl2019}, and ride-pooling~\cite{PaparellaPedrosoEtAl2023}.
For electric fleets, the operational problem encompasses the charging schedule.
To this end, fast solution algorithms based on acyclic graphs have been proposed by Yao et al.~\cite{YaoChenEtAl2021b}. In~\cite{RossiIglesiasEtAl2018b,EstandiaSchifferEtAl2021}, the authors accounted for the coupling with the power grid via network flow models. 
Regarding E-AMoD, multi-layer network flow models inspired by~\cite{RossiIglesiasEtAl2018b} have been recently leveraged to optimize the charging station siting and sizing jointly with the operation of the fleet~\cite{LukeSalazarEtAl2021,PaparellaChauhanEtAl2023}.
Nevertheless, the majority of these papers do not focus on vehicle-level design aspects, if not via parametric studies of the vehicles. Usually, the vehicle is assumed to be given because it allows to pre-define the mass, energy consumption and autonomy range of the single vehicle.

The design of AMoD systems has been investigated with methods ranging from Directed Acyclic Graphs (DAGs)~\cite{PaparellaHofmanEtAl2022}, to fluidic models~\cite{ZardiniLanzettiEtAl2020b} which are treated as a linear time invariant problems. 
Wallar et al.~\cite{Wallar2019,Wallar_2019} devised an algorithm to capture vehicles with a different seat-capacity and optimize their number and operation for a ride-sharing AMoD environment, but without considering an electric fleet that needs to be recharged. In~\cite{HoffAnderssonEtAl2010b,BaltussenGouthamEtAl2023}, the authors investigated the VRP with time windows and heterogeneous fleet composition, but they did it by selecting from a pre-existing set of vehicles so that they would minimize a given objective.

In conclusion, to the best of the authors' knowledge, there is a lack of an optimization framework that simultaneously optimizes the number of vehicles, their battery size, and the operation of the fleet. 

\emph{Statement of Contributions:}
The contribution of this paper is threefold. 
First, we propose an optimization framework for E-AMoD systems based on DAGs. The optimization problem includes vehicle-level design variables of both the single vehicle unit and of the entire fleet itself.
Second, to overcome tractability issues of large problem instances, we devise and analyze a method based on solving multiple randomly sampled sub-problems to draw the probability distribution of the design solution. This allows to find a slightly more conservative solution, in line with the design objective of the optimization problem. 
Third, we present our results on a real-world case study for Manhattan, NYC, USA, and we show the trade-offs between number of vehicles, battery size and electricity cost. 
A preliminary version of this paper was presented at the 2022 IEEE Conference on Decision and Control~\cite{PaparellaHofmanEtAl2022}.
In this extended version, we carry out a broader literature review, include battery degradation in the model, analyze the quality of the solution obtained from the randomly sampled sub-problems, and provide a heuristic solution to increase their solvable size. Moreover, we conduct a sensitivity analysis on the battery capacity of the fleet to better capture the trade-offs between the design variables. 

\emph{Organization:}
The rest of the paper is organized as follows: Section~\ref{Sec:GenFrame} introduces the optimization framework and its underlying assumptions. Section~\ref{sec:sampleds} provides a solution approach. The case study for Manhattan, NYC is discussed in Section~\ref{Sec:Res}. In the final Section~\ref{Sec:Concl}, we summarize the work, offer a discussion and suggest avenues for future research.

%% file: Sections/GeneralFramework.tex
\section{Optimization Problem Formulation}\label{Sec:GenFrame}
This section formulates the optimal vehicle assignment and charge scheduling problem leveraging DAGs. Thereafter, we include the formulation of the objective function, constraints and variables, capturing the trade-off between number of vehicles, battery capacity of the single unit, cost to operate the fleet, and revenues generated by serving travel requests.

\subsection{Road Network}
We model the transportation system as a directed graph \mbox{$\mathcal{G'} = (\mathcal{V'},\mathcal{A'})$}, where the set of arcs $\mathcal{A'}$ represents road links, the set of vertices $\mathcal{V'}$ contains intersections.
We also indicate $D_{mn}$ and $T_{mn}$ as the distance and travel time, respectively, of road segments between road intersections $(m,n) \in\mathcal{A'}$. We denote a set of travel requests by $\mathcal{I}$ with $ i \in\mathcal{I} := \{1,2,...,I\}$ the set of transportation requests. In order to model the demanded trips, let the triple $r_i = (o_i,d_i,t_i^\mathrm{start})$ denote a requested trip, where $t_i^\mathrm{start}$ is the requested pick-up time, whilst $o_i,d_i \in \mathcal{V'}$ are the origin and destination nodes of request $i$, respectively.
In the area under consideration there are $C$ charging stations, whereby each station $c \in\mathcal{C} :=\{1,2...,C\}$ is located at vertex $n_c \in \cV'$.
For each arc in $\mathcal{A'}$ we assume the driving pattern and the travel time to be fixed and known in advance for each time of the day. 
Finally, we assume that vehicles drive through the fastest path when traveling from one location to another. Yet other criteria can be readily implemented to predefine the paths between locations.

\subsection{Directed Acyclic Graph}
In order to study the fleet design and operation problem in a mathematically tractable fashion, we construct an acyclic directed graph $\mathcal{G}$, similar to~\cite{LamLeungEtAl2016}. 
Each arc in $\cG$ represents a specific pre-computed route to be taken in the original graph $\cG'$ to travel from the destination of a  request to the origin of the next one. To include depots, we define an extended set of requests $\mathcal{I}^+:=\{0,1,2...,I,I+1\}$ where the depots are the first and last requests that have to be served so that vehicles start and conclude their schedules in a depot. In other words, graph $\mathcal{G'}$ describes the geography of the road network, where the collection of arcs connecting two locations represents a path. The DAG $\mathcal{G}$ represents the sequential order of requests that are served. 
This way we can capture the transitions between requests: Each arc $(i,j) \in \mathcal{A}$ represents the transition from the destination of $i$, $d_i$, to the origin of $j$, $o_j$, and it is characterized by the travel time and distance $t^\mathrm{fp}_{ij}$ and $d^\mathrm{fp}_{ij}$, respectively, of the fastest path.
If $i=j$, then $t^\mathrm{fp}_{ii}$ is merely the fastest time to serve $i$. Given the set of $K\in\sN$ vehicles $\cK :=\{1,2...,K\}$, to capture whether vehicle $k\in\cK$ serves request $i$ and then request $j$, we set the binary tensor $X_{ij}^k=1$ and to $0$ otherwise. Last, when taking into account also the state of energy of the vehicles, the DAG represented in an energy-time space, is defined as E-DAG.
If between the two requests vehicle $k$ charges its battery at charging station $c$, we set the binary tensor $S_{ijc}^k=1$ and $0$ otherwise; we also quantify the amount of battery charged with the non-negative-valued tensor $C_{ijc}^k$. Nevertheless, it is possible to also account for vehicle-to-grid (V2G) activities by relaxing tensor $C$ to negative values, as in~\cite{PaparellaHofmanEtAL2023}. 

\subsection{Example}\label{sec:example}
In this section, in Fig.~\ref{fig:DAG}, we demonstrate, in a reader-friendly fashion, the case of three requests to be served, $\cI = [1,2,3]$. We extend the set of requests $\cI^+ = [0,1,2,3,4]$, and we identify the depot with nodes $0$ and $4$, meaning that the vehicle starts and ends the trip in a pre-defined location. If vehicle $1$ first serves request $1$ and then request $2$, then $X_{12}^1=1$.
It means that the vehicle serves in order the two requests and then idles for the rest of the time.
During the final transition between request 3 and 4 (the depot), if vehicle $1$ is charged at charging station $1$ after serving request $3$ and before arriving to request $4$, then $X^1_{34}=1$ and $S_{231}^1=1$. To indicate that charging occurs during the transition, the arc in the figure is depicted in green. The amount of energy recharged in $S_{231}^1=1$ is defined as $C_{231}^1$. We highlight that with this formulation, because of the absence of time dependency, temporal and sequential information about the actions inside a transition are lost. For example, in $X_{23}^1=1$ there might be a time interval when the vehicle idles. This can occur before traveling to the charging station, during the stay in the station, or at the depot. The information about the duration of the idling time is retrievable, but it cannot be allocated to a precise time-slot without further assumptions on the sequentiality of tasks.

\subsection{Objective Function}
In this paper, we set as optimization objective the maximization of profit for the fleet operator. We identify three main terms that play a key role: the fixed costs to purchase the fleet of vehicles, the variable costs to operate the fleet, and the revenues generated by serving the travel requests.
We define $p_0 ^k$ as vehicle $k$'s amortized cost---the process of gradually writing off the initial cost over the lifetime of the vehicle $\tau_\mathrm{v}$---as an affine function of its battery energy capacity $E_\mathrm{b}^k$:
\begin{equation}\label{eq:p0}
	p_0 ^k  =    \frac{p_\mathrm{v} \cdot b_\mathrm{v}^k+ p_\mathrm{b} \cdot E_\mathrm{b}^k}{\tau_\mathrm{v}}\;\;\;\;\forall k\in \cK.
\end{equation}
Note that $E_\mathrm{b}^k$ can be different for each vehicle, enabling a heterogeneous fleet. Without loss of generality, for a fleet to be homogeneous, each battery capacity $E_\mathrm{b}^k$ can be set equal to the others. 
The price to purchase the entire vehicle excluding the battery is defined as $p_\mathrm{v}$, while $p_\mathrm{b}$ indicates the price per unit energy of battery capacity. Note that this last term can include not only the price of the battery itself, but also the costs that depend on the battery, i.e., a very large battery requires a heavier and larger chassis. The variable $p_0^k$ captures the daily fixed cost per vehicle, that can be conveniently adjusted to include additional factors like insurance, storage and maintenance. The  binary variable $b_\mathrm{v}^k \in \{0,1\}$ indicates if vehicle $k$ is used (and purchased) or not. 
We then define the operational cost as the price paid to charge the whole fleet for one day. The electricity price, indicated by $p_\mathrm{el}$, is the price per unit energy, which is set as a constant. However, it can conveniently be set as a function of time, i.e., as a function of transition $ij$. The total amount of energy charged on the day by the whole fleet is $C^\mathrm{tot}$, defined as
\begin{equation}
	C^\mathrm{tot}= \sum_{i,j \in \cI^+}\sum_{c \in \cC} \sum_{k \in \cK} C_{ijc}^k.
\end{equation}  
Last, we define $p_i$ as the revenue generated by serving the $i$-th request, which can be given or as in our case, computed. 
We also include the choice of serving request $i$ by introducing $b_\mathrm{r}^i$, the binary variable indicating if request $i$ was served or not.

To conclude, we define the (negative) profit objective function as a linear combination of the previously explained terms as follows:
\begin{equation}\label{eq:obj}
J= \sum_{k \in \cK} p_0^k + p_\mathrm{el} \cdot C^\mathrm{tot}  - \sum_{i \in \cI^+} b_\mathrm{r}^i \cdot p_i.
\end{equation}

\subsection{Operational Constraints}\label{sec:oper}
We define the transition matrix $X \in \{0,1\}^{\abs{\cI^+} \times \abs{\cI^+} \times K}$, where $X_{ij}^k=1$ if vehicle $k$ serves demand $i$ and then demand $j$, and zero otherwise. We also introduce the tensor $S \in \{0,1\}^{\abs{\cI^+} \times \abs{\cI^+} \times K \times C}$ to account for charging activities. If vehicle $k$ in-between $d_i$ and $o_j$ goes to charging station $c$, $S_{ijc}^k$ is set to one, otherwise it is set to zero.
In the same way, we quantify the amount of energy charged with element $C_{ijc}^k$ of the corresponding charging tensor $C \in \sR_{+}^{\abs{\cI^+} \times \abs{\cI^+} \times K \times C}$.
However, because of time constraints, not all transitions are feasible. To determine whether a transition is feasible or not, we define the available time between transitions as \mbox{$t_\mathrm{ava}: = t_j - t_i + t_{ii}^\mathrm{fp}$}, that represents the time between $d_i$ and $o_j$. 
Travel, deviation and charging times are efficiently pre-computed via standard shortest path algorithms, so we can pre-compute which transitions are feasible and directly eliminate such unfeasible variables, i.e., compute upper bounds. It follows that
\begin{equation}
	X_{ij}^k\leq 
	\begin{cases}
		1 & \text{if} \; t_{ij}^{\mathrm{fp}} \leq t^\mathrm{ava}_{ij}\\
		0 & \text{otherwise}
	\end{cases}  \;\;\forall i,j\in\cI^+,\;\;\forall k\in\cK.
\end{equation}
Via this upper bound, we obtain a triangular adjacency matrix, from which derives the DAG formulation, as described in~\cite{LamLeungEtAl2016,YaoChenEtAl2021b}.
In the same way, if there is enough time available, a deviation to charging station $c$ within transition $ij$ is feasible:
\begin{equation}
	{S}_{ijc}^k \leq 
	\begin{cases}
		1 & \text{if} \; t_{ij}^\mathrm{fp} +  \Delta T_{ijc}^\mathrm{go2S} \leq t^\mathrm{ava}_{ij}\\
		0 & \text{otherwise}
	\end{cases}   \;\;\forall i,j\in\cI^+,\;\;\forall k\in\cK.
\end{equation}
Finally, we write the upper bound of the amount of energy that can be charged at station $c$ as
\begin{equation}\label{eq:Cmax}
	{C}_{ijc}^k \leq
	\begin{cases}
		\hat{C}_{ijc}^k & \text{if} \; t_{ij}^\mathrm{fp} +  \Delta T_{ijc}^\mathrm{go2S} \leq t^\mathrm{ava}_{ij}\\
		0 & \text{otherwise}
	\end{cases}  
	\;\;\forall i,j\in\cI^+,\;\;\forall k\in\cK,
\end{equation}
where $\hat{C}_{ijc}^k=\mathrm{min}[( t^\mathrm{ava}_{ij} -t_{ij}^\mathrm{fp} - \Delta T_{ijc}^\mathrm{go2S})\cdot P_\mathrm{ch},E_\mathrm{b}^\mathrm{max}] $ is the upper bound of the energy that can potentially be charged if all the available time left were used to charge. Note that we introduce $\mathrm{min}(\cdot)$ because the maximum amount charged cannot be greater than the capacity of the battery size itself, with $E_\mathrm{b}^\mathrm{max}$ being an upper bound of the battery capacity of the fleet.
We then enforce that vehicle $k$ can only charge an amount $C_{ijc}^k \geq 0$ if $S_{ijc}^k= 1$, which, in turn, can happen only if transition $ij$ is performed:
\begin{equation}\label{eq:SOnlyIfX}
\sum_{c\in \cC}S_{ijc}^k \leq  X_{ij}^k \;\;\;\;\forall i,j\in \cI^+,\;\forall k\in \cK,
\end{equation}
\begin{equation}\label{COnlyIfS}
C^k_{ijc} \leq \hat{C}_{ijc}^k \cdot S_{ijc}^k\;\;\;\;\forall i,j\in \cI^+,\;\forall c\in \cC,\;\forall k\in \cK.
\end{equation}
We define two parameters $f$ and $l$, so that $f_j^k,l_j^k=0 \;\forall i,j \in \cI$ and $f_0^k=l_{I+1}^k=1\;\forall k\in\cK$, to initialize and finalize the vehicles in a specific location, the depot.
Then, we enforce that each request can be served once, at most.
In other words, before and after a served request, there can only be a previous and a subsequent  request, at most, that can be served, which is expressed by
\begin{equation}\label{eq:maxonce1}
   \sum_{i\in \cI^+,k\in \cK}  X_{ij}^k+ \sum_{k\in \cK} f_{j}^k \leq 1\;\;\;\;\forall j\in \cI^+,
\end{equation} 
\begin{equation}\label{eq:maxonce2}
   \sum_{j\in \cI^+,k\in \cK}  X_{ij}^k+ \sum_{k\in \cK} l_{i}^k \leq 1 \;\;\;\;\forall i\in \cI^+.
\end{equation}  
Finally, we ensure continuity of the schedule of each vehicle, meaning that if vehicle $k$ serves request $i$, transition $ij$ can only be effected by the same vehicle $k$,
\begin{equation}\label{eq:conti}
\sum_{i\in \cI^+} X_{ij}^k - \sum_{l\in \cI^+} X_{jl}^k = f_j^k+l_j^k\;\;\;\;\forall j\in \cI^+,\;\forall k\in \cK.
\end{equation}
We highlight that in~\eqref{eq:maxonce1}--\eqref{eq:conti} the terms $f_j^k$and $l_j^k$ are always null for $i,j \in \cI$. This is not the case if \mbox{$i,j \in \{0,I+1\}$}, i.e., at the beginning or end of the schedule, where the first and last transitions are initialized or finalized in a depot.
    
\subsection{Energy Constraints}\label{sec:en}
In this section we introduce the energy balance of each vehicle $k$ for every node of the DAG. We express the state of battery charge of vehicle $k$ on a given node $j$ as $e^k_{j}$, that represents the energy at the end of trip $j$:
\begin{align}
e^k_{j} =  e^k_{i} - d^\mathrm{fp}_{jj}\cdot& \Delta e^k - E_{ij}^k + \sum_{c\in \cC} C_{ijc}^k\nonumber\\
&\forall i,j \in \cI^+,\forall k \in \cK |X_{ij}^k=1,\label{eq:EbalJ}
\end{align}
with $\Delta e^k$ being the consumption per unit distance, $d^\mathrm{fp}_{jj}$ the distance of the fastest path between $o_j$ and $d_j$.
The energy at the end of trip $j$ is equal to the energy at the end of the previous trip, minus the energy to serve it, minus the transition energy $E_{ij}^k$, plus the charged energy, $C_{ijc}^k$, if any. We recall that~\eqref{eq:EbalJ} is valid only if $X_{ij}^k = 1$.
We re-write~\eqref{eq:EbalJ} with the big M formulation to take it into account, where $M$ is a sufficiently large number~\cite{GrivaNashEtAl2019}:
\begin{equation}\label{eq:energy1}
\begin{split}
	e^k_{j} \geq  e^k_{i}- d^\mathrm{fp}_{jj}\cdot \Delta e^k - E_{ij}^k + &\sum_{c\in \cC}C_{ijc}^k - M \cdot (1-X_{ij}^k) \\
	&\forall i,j\in \cI^+, \forall k\in \cK,
\end{split}
\end{equation}
\begin{equation}\label{eq:energy2}
\begin{split}
	e^k_{j} \leq  e^k_{i} - d^\mathrm{fp}_{jj}\cdot \Delta e^k - E_{ij}^k + &\sum_{c\in \cC}C_{ijc}^k + M \cdot (1-X_{ij}^k)\\
	&\forall i,j\in \cI^+,\forall k\in \cK.
\end{split}
\end{equation}
The energy to transition from $d_i$ to $o_j$, $E_{ij}^k$, is defined by
\begin{equation}\label{eq:EbalTrans}
E^k_{ij} = 
\begin{cases}
\begin{split}
d^\mathrm{fp}_{ij}\cdot &\Delta e^k \\ \; & \; \forall i,j \in \cI^+,\forall k \in\cK |\sum_{c\in \cC} S_{ijc}^k=0,
\end{split} 
\\
\begin{split}
(d^\mathrm{fp}_{ij} &+ \Delta d_{ijc}^\mathrm{go2S} ) \cdot \Delta e^k  \\
&\forall i,j \in \cI^+,\forall c \in\cC,\forall k \in\cK |S_{ijc}^k=1.
\end{split}
\end{cases}
\end{equation}
If a detour to charging station $c$ occurs, $\Delta d_{ijc}^\mathrm{go2S}$ is the additional distance traveled to pass through it. We reformulate \eqref{eq:EbalTrans} with the big M formulation:
\begin{equation}\label{eq:EbalTrans1}
\begin{split}
E_{ij}^k \geq \Delta e^k  \cdot   d^\mathrm{fp}_{ij} \;\;& \forall i,j \in \cI^+,\forall k \in\cK,
\end{split}
\end{equation}
\begin{equation}\label{eq:EbalTrans2}
\begin{split}
    E_{ij}^k \leq \Delta e^k  \cdot d^\mathrm{fp}_{ij} + M \cdot  &\sum_{c\in \cC} S_{ijc}^k \\& \forall i,j \in \cI^+,\forall k \in\cK,
\end{split}
\end{equation}
\begin{equation}\label{eq:EbalTrans3}
\begin{split}
   E_{ij}^k \geq \Delta e^k \cdot (d^\mathrm{fp}_{ij}&+\Delta d^\mathrm{go2S}_{ijc}) - M \cdot (1-S_{ijc}^k) \\ 
   & \forall i,j \in \cI^+,\forall k \in\cK,\forall c \in\cC,
\end{split}
\end{equation}
\begin{equation}\label{eq:EbalTrans4}
\begin{split}
    E_{ij}^k \leq \Delta e^k \cdot (d^\mathrm{fp}_{ij}&+\Delta d_{ijc}^\mathrm{go2S}) + M \cdot (1-S^k_{ijc})\\
    & \forall i,j \in \cI^+,\forall k \in\cK,\forall c \in\cC.
\end{split}
\end{equation}
Eq.\eqref{eq:EbalTrans1} is always active, \eqref{eq:EbalTrans2} becomes inactive if \mbox{$\sum_c S_{ijc}^k=0$}, \eqref{eq:EbalTrans3} and \eqref{eq:EbalTrans4} become active if \mbox{$\sum_c S_{ijc}^k=1$}.

Assuming a fairly constant battery-to-wheels efficiency we formulate the vehicle's consumption per unit distance as an affine function of its mass~\cite[Ch.~2]{GuzzellaSciarretta2007}.
Since the mass is, in turn, an affine function of the battery size, the vehicle consumption per unit distance $\Delta e^k$ is
\begin{equation}\label{eq:consumes}
	\Delta e^k = \Delta e_0 + \Delta e_\mathrm{b}\cdot E_\mathrm{b}^k \;\;\;\;\forall k\in \cK.
\end{equation}
The base vehicle consumption is $\Delta e_0$, whereas $\Delta e_\mathrm{b}$ is a linear term.
Thereafter, the battery size must always be larger than the energy stored in it. Last, the energy stored must always be greater than zero:
\begin{equation}
  E_ \mathrm{b}^k \geq e^k_{j} \geq 0 \;\;\forall j \in \cI^+,\;\forall k\in\cK\\.
\end{equation}

\subsection{Number of Vehicles} 
In this section we introduce constraints related to the number of vehicles that can used used in the fleet.
First, we define the binary variable $b_\mathrm{v}^k=\{0,1\}$ to indicate whether vehicle $k$ is being used or not. Note that this formulation can be generalized by extending the domain of $b_\mathrm{v}^k$ to integer values. In this case it would be straightforward to take into account multiple vehicles with different characteristics. 
As previously explained at the beginning of Section~\ref{Sec:GenFrame}, depots are modeled as the first and last requests to be served and are included in $\cI^+$.
If a vehicle is not used, it stays in the depot, meaning that the only transition it performs is $X^k_{0,I+1}=1$. It follows that
\begin{equation}
\sum_{i,j\in \cI^+}X_{ij}^k \leq 1-M \cdot b_\mathrm{v}^k \;\;\;\; \forall k \in \cK,
\end{equation}
\begin{equation}
	E_\mathrm{b}^k \leq M \cdot b_\mathrm{v}^k \;\;\;\; \forall k \in \cK.
\end{equation}
Then, we enforce the energy of each vehicle at the beginning and end of the schedule to be the same, which, in turn, can ba set equal to an inizialization parameter, $E_\mathrm{b}^0$,
\begin{equation}
	e^k_{0} = e^k_{I+1} = b_\mathrm{v}^k \cdot E_\mathrm{b}^0\;\;\forall k\in\cK.
\end{equation}
If the vehicle is not being used, $b_\mathrm{v}^k = 0$ and the initial and final battery state is equal to \unit[0]{kWh}. Note that in this way it is possible to avoid the non linear term $ b_\mathrm{v}^k \cdot E_\mathrm{b}^k$ in~\eqref{eq:obj}.
Finally, to decrease the number of multiple solutions, without loss of generality, we use vehicles sequentially:
\begin{equation}\label{eq:vehiprior}
  b_\mathrm{v}^{k+1} \leq b_\mathrm{v}^k\;\;\;\; \forall k \in \cK.
\end{equation}

\subsection{Problem Formulation}
To summarize, we formulate the maximum-profit design and operation problem for an E-AMoD fleet as follows:

\begin{prob}[Joint Design and Operation Optimization]\label{prob:main}
	Given a set of transportation requests $\{r_i\}_{i \in \cI}$ and a set of charging stations $\cC$, the number of vehicles, their battery size, and their operations maximizing the total profit result from
	\begin{equation*}
		\begin{aligned}
			\min\; &J \\
			\mathrm{s.t.}\; &\eqref{eq:p0}-\eqref{eq:conti},\; \eqref{eq:energy1}-\eqref{eq:energy2},\;\eqref{eq:EbalTrans1}-\eqref{eq:vehiprior}. 
		\end{aligned}
	\end{equation*}
\end{prob}
Problem~\ref{prob:main} is a mixed integer linear program that can be solved with global optimality guarantees by off-the-shelf optimization algorithms.

\subsection{Discussion}\label{Sec:discussion}
A few comments are in order.
First, we consider travel times on the road digraph $\cG'$ to be given. This assumption is in order for a small fleet as the one under consideration, whose routing strategies do not significantly impact travel time and hence overall traffic. This way, also varying levels of exogenous traffic during the course of the day can be captured by simply including time-dependent traffic data and adjusting fastest path time and distance accordingly.
Second, we assume the charging stations to always be available. We leave the inclusion of constraints to avoid potentially conflicting charging activities by multiple vehicles to future research. Moreover, the charging power  $P_\mathrm{ch}$ is a parameter that can be conveniently adapted to a specific charging infrastructure.
Third, considering design aspects, the solution of Problem~\ref{prob:main} is deterministic. Optimizing the fleet for a specific scenario may render its design not feasible for another one. 
This problem can be addressed by either a robust optimization approach or, as we do in this paper, by solving the problem for multiple sampled scenarios---i.e., solving Problem~\ref{prob:main} with a subset of travel requests---as already done in the literature~\cite{PhamLeymanEtAl2018,Lopez-IbanezBlum2010,JuillePollack1998}.
In this way, it is possible to draw the probability distribution of the design solution. The obtained solution is sub-optimal compared to directly solving the whole problem, as will be shown in Section~\ref{sec:sampleds}. However, this leads to a more conservative design solution that guarantees more robustness to different scenarios w.r.t.\ solving a larger instance of Problem~\ref{prob:main}.

%% file: Sections/Sampling.tex
\section{Solution Approach}\label{sec:sampleds}
Problem~\ref{prob:main} is a MILP that can be solved with commercial solvers. However, due to its combinatorial nature, solving it for a large number of travel requests might lead to computational intractability. Moreover, as previously discussed in Section~\ref{Sec:discussion}, the solution of the problem is deterministic. To mitigate both the shortcomings highlighted above, we devise the method that we explain in the following.

\subsection{Randomly Sampled Sub-problems}
We infer a conservative solution of the design problem by solving $m$ times a set of sub-problems, also called scenarios, with $n$ travel requests each, where each sub-problem set of travel requests is randomly sampled from the original set.

Thereafter, we recover a distribution of the solution of the original problem as the aggregate of the solutions of all the sub-problems.
Notably, the smaller the scenario and the number of demands, the more difficult it is to match demands with vehicles in an efficient manner.
In contrast, the larger the scenario, the easier it will be to coordinate the vehicles due to the so-called Mohring effects~\cite{FielbaumTirachiniEtAl2021}, ultimately leading to a smaller fleet.
Therefore, the smaller the sub-problem size, the more conservative the resulting aggregate solution will be in terms of estimated rejection rate, number of vehicles and achieved objective, as will be quantitatively shown in the Results Section~\ref{sec:size}. 


\subsection{Heuristic Sub-problem Reduction}\label{sec:omniscient}
In the previous section, we explained that it is possible to infer a more conservative solution of the whole problem by solving many randomly sampled sub-problems and aggregating their solutions. We also highlighted that, the smaller the sub-problem, the more conservative the aggregate solution is. For this reason, the size of each sub-problem should be maximized while taking into account the computational complexity. Given the combinatorial nature of the problem, optimizing over many travel requests would mean having an extremely large amount of variables that would, in turn, quickly lead to an intractable problem. Moreover, compared to the standard assignment problem, the framework in this paper cannot be solved neither by Hungarian algorithm~\cite{Munkres1957}, nor by relaxing the integer variables, because the energy constraints shown in Section~\ref{sec:en} are not entirely unimodular.
Against this backdrop, and similar to~\cite{VazifehSantiEtAl2018}, we shrink the feasible domain, removing very improbable solutions.
In particular, for two travel requests $i$ and $j$ so that $t_{ij}^\mathrm{fp} \leq t_{j}^\mathrm{start}-t_i^\mathrm{end}$, the upper bound of $X_{ij}^k$ is equal to $1$, even if the two requests are spatial-temporally very distant one from the other.
On the one hand, by not restricting the space of variables, we would not lose guarantees of global optimality for the sub-problem. On the other hand, the larger the instance of the sampled problem, the better the quality of the aggregate solution w.r.t. the one the full size problem. 
Thus, we restrict the upper bounds of tensors $X,S,C$ so that each transition $ij$ can only happen if the idling time and the rebalancing distance are below a given amount $\bar{t}^{\mathrm{idl}}_{ijc}$.
We highlight that, whilst in small problem instances such a constraint could significantly affect the solution, for large problem instances it can be easily respected and can be leveraged to efficiently reduce the feasible domain.
Formally, we define $t^{\mathrm{idl}}_{ijc}$ as the idling time of transition $ij$ passing through charging station $c$ and restrict it as
 \begin{equation}\label{eq:excess time}
 	t^{\mathrm{idl}}_{ijc} = t_j^{\mathrm{start}}- t_i^{\mathrm{end}} - \Delta T^\mathrm{go2S}_{ijc} - t_{ij}^{\mathrm{fp}} - \frac{E_{\mathrm{day}}}{P_{\mathrm{ch}}},
 \end{equation}
 \begin{equation}\label{eq:ineqHeur}
 	t^{\mathrm{idl}}_{ijc}\leq \bar{t}^{\mathrm{idl}}_{ijc} \qquad \quad \forall i,j \in \mathcal{I}, \forall k \in \mathcal{K}, \forall c \in \mathcal{C}.
 \end{equation}
 If~\eqref{eq:ineqHeur} is not satisfied, the corresponding upper bounds of the elements of $X^k_{ij},S^k_{ijc},C^k_{ijc}$ are set to $0$. 
We define:
 \begin{equation}
 	\begin{split}
 		&\Delta T^\mathrm{go2S}_{ijc} =\Delta \hat{T}^\mathrm{go2S}_{ijc} \quad \mathrm{if} \quad  \Delta \hat{T}^\mathrm{go2S}_{ijc} \leq \Delta \bar{T}^\mathrm{go2S}_{ijc}, \\
 		&\Delta T^\mathrm{go2S}_{ijc} = -\infty \qquad \; \mathrm{otherwise}, \\
 		&t_{ij}^{\mathrm{fp}} =\hat{t}_{ij}^{\mathrm{fp}} \qquad \quad \quad \quad \;\mathrm{if} \qquad  \hat{t}_{ij}^{\mathrm{fp}} \leq \bar{t}_{ij}^{\mathrm{fp}}, \\
 		&t_{ij}^{\mathrm{fp}} = -\infty \qquad \qquad \;\; \mathrm{otherwise},
 	\end{split}
 \end{equation}
where the barred terms on the right side are parameters subject to tuning. They are thresholds up to which the idling time is set to infinity and~\eqref{eq:ineqHeur} does not hold.
The last term represents the maximum time a vehicle can spend in a charging station, which still allows for far-apart transitions that would occur because the vehicle needs time to charge.
Overall, by applying this heuristic, it is possible to decrease the feasible domain by approximately $80\%$, improving significantly the tractability and scalability of the problem so that larger instances can be solved.

%% file: Sections/Results.tex
\section{Results}\label{Sec:Res}
This section showcases our framework for Manhattan, NYC.
Specifically, we use the road network shown in Fig.~\ref{fig:DAG},  consisting of 357 nodes and 1006 links, which was constructed using a version based on OpenStreetMaps~\cite{HaklayWeber2008}.
The travel requests were supplied by the Taxicab \& Livery Passenger Enhancement Programs to the NYC Taxi and Limousine Commission. The data set is built using historical data of taxi rides that occurred in March 2018. 
The values of all the parameters used are collected  in Table~\ref{table}, together with their sources. Following~\cite{UberPrice}, we express it as an affine function with respect to time and distance required to serve the travel request:
\begin{equation}
p_{i} = \alpha + \beta \cdot d^\mathrm{fp}_{ii} + \gamma \cdot t^\mathrm{fp}_{ii}\;\;\;\forall i\in \cI, 
\end{equation}
with $\alpha$ equal to the base fare, $\beta$ to the cost per unit distance and $ \gamma$ to the cost per unit time. The number of cycles to end-of-life of the battery of the vehicles, $\tau_\mathrm{v}^\mathrm{cycle}$, is further explained in Section~\ref{sec:lifetime}. 
 
Following the method explained in Section~\ref{sec:sampleds}, we solve multiple randomly sampled sub-problems, obtaining a discretized sub-optimal distribution of the original solution of the full problem. 
Based on trial-and-error, we set $\Delta \bar{T}^\mathrm{go2S}_{ijc},\bar{t}_{ij}^{\mathrm{fp}},	\bar{t}^{\mathrm{idl}}_{ijc}= \unit[0.1]{h}$, $E_{\mathrm{day}}=\unit[30]{kWh}$.
Finally, we assume that the E-AMoD operator has 15 privately owned charging facility infrastructures, evenly distributed in the area of interest as shown in Fig.~\ref{fig:DAG}, that is a sufficient number to avoid large distance recharging trips~\cite{PaparellaChauhanEtAl2023}. 
In future research, will also address the issue of jointly optimizing, taking into account the siting and sizing of the charging infrastructure. This can readily be implemented by adding an integer variable for each station and changing the objective function accordingly.
We parse and solve Problem~\ref{prob:main}, using Yalmip~\cite{Loefberg2004} and Gurobi~9.1~\cite{GurobiOptimization2021}. 

\begin{table}[!t]
	\caption{Values of Parameters}
	\label{table}
	\begin{center}
		\begin{tabular}{|c||c||c||c|}
			\midrule
			Parameter & Value & Unit & Reference\\
			\midrule
			$\tau^\mathrm{cycle}_\mathrm{v}$ & $2500$ & cycles &\cite{Burns2013}\\
			\midrule
			$p_\mathrm{v}$ & $8000$ & €&\cite{BOSCH201876}\\
			\midrule
			$p_\mathrm{batt}$ & $700$ & €/kWh&\cite{BloombergBattery}\\
			\midrule
			$p_\mathrm{el}$ & $0.30$ & €/kWh&\cite{LaborStatistics}\\
			\midrule
			$P_\mathrm{ch}$ & $6$ & kW&\cite{ChargeHub.com}\\
			\midrule
			$\Delta e_0$ & $0.09$ & kWh/km&\cite{GuzzellaSciarretta2007}\\
			\midrule
			$\Delta e_\mathrm{b}$ & $0.0025$ & 1/km&\cite{GuzzellaSciarretta2007}\\
			\midrule
			$\alpha$ & $2.55$ & €&\cite{UberPrice}\\
			\midrule
			$\beta$ & $1.5$ & €/km&\cite{UberPrice}\\
			\midrule
			$\gamma$ & $0.35$ & €/min&\cite{UberPrice}\\
			\midrule
		\end{tabular}
	\end{center}
\end{table}

\subsection{Numerical Experiments on the Problem Size}\label{sec:size}
\begin{figure}[t!]
		\centering
		\includegraphics[trim={0cm 0 30 5},clip,width=\linewidth]{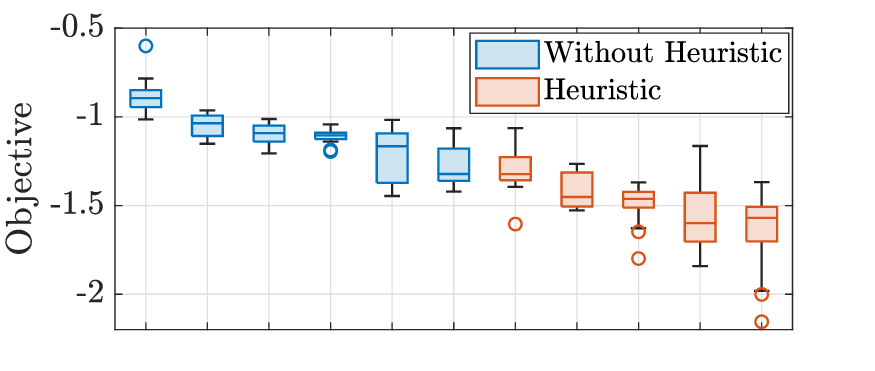}
		\includegraphics[trim={0cm 0 30 5},clip,width=\linewidth]{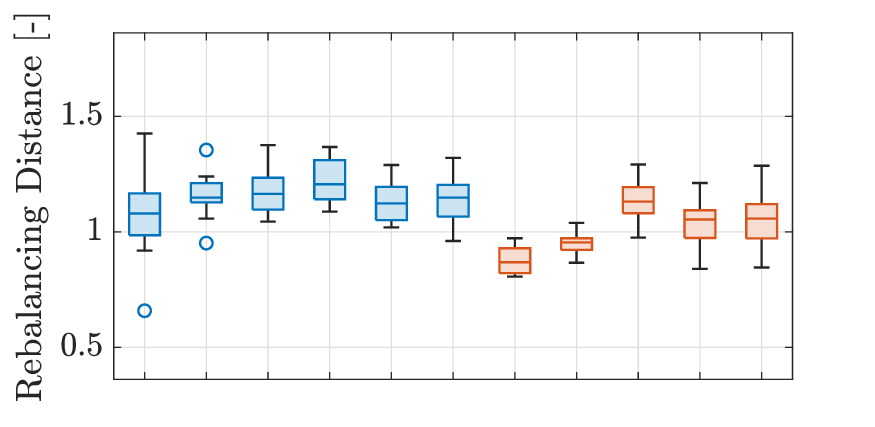}
		\includegraphics[trim={0cm 0 30 5},clip,width=\linewidth]{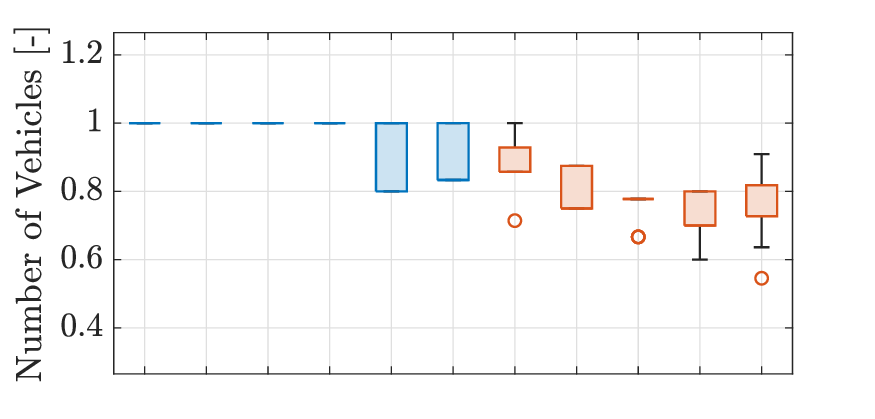}
		\includegraphics[trim={0cm 0 30 5},clip,width=\linewidth]{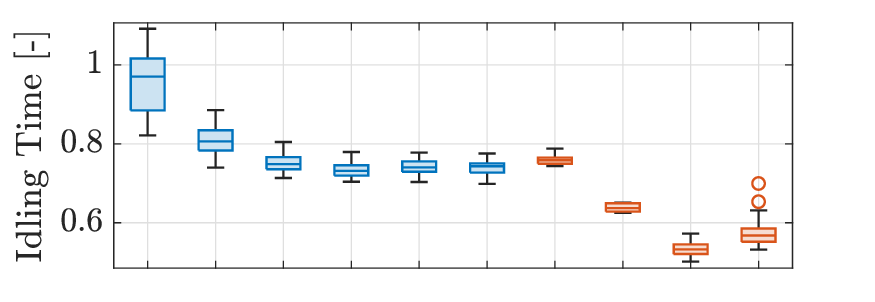}
		\includegraphics[trim={0cm 0 30 5},clip,width=\linewidth]{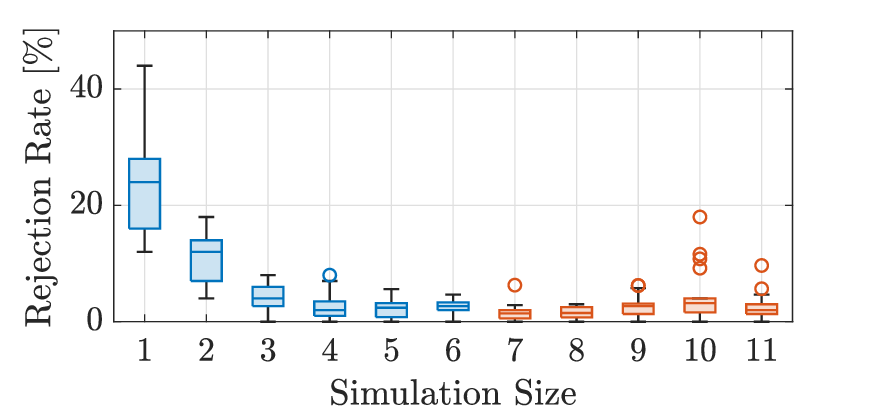}
		\caption{The figure shows the normalized objective per vehicle and the rejection rate as a function of the size of the simulation. Simulation size 1 is equal to 1 vehicle and 25 travel requests. The battery size of the vehicles is fixed to \mbox{$E_\mathrm{b}^k = \unit[20]{kWh} \;\; \forall k \in \cK$}. It can be noted that  the larger the number of vehicles, the lower the rejection rate and the lower the objective function, the higher the profit.} 
		\label{fig:theo}
		
	\end{figure}

To quantitatively analyze the impact of the size of the sub-problems on the solution, we conduct a scan of the solution distribution w.r.t. it, as shown in Fig.~\ref{fig:theo}. The heuristic approach to increase the solvable size of the sub-problems is explained in Section~\ref{sec:omniscient}. All the results presented are normalized. 
The simulation clearly depicts that the smaller the instances, the more sub-optimal the solution. In fact, the larger the problem instance, the larger the profit---the smaller the objective function---, and the lower the number of vehicles required. Note that for a sufficiently large scenario, the coordination between vehicles allows to reduce their number, as indicated by the lower value of the normalized number of vehicles. Nevertheless, the computation time increases exponentially. We recall that: For very small scenarios the heuristic method in Sec.~\ref{sec:omniscient} is unsuitable due to the overall low number of requests; For large scenarios the problem can only be solved with the heuristic method. Interestingly, the phenomenon observed in Fig.~\ref{fig:theo} recalls the Better-matching and Mohring effects~\cite{FielbaumTirachiniEtAl2021}: the higher the number of people requesting for a mobility service, the higher the performance of the system. Thus, the larger the instance, the better. 
However, depending on the computational and time resources, and on the quality of the solution required, it is possible to compute a sub-optimal solution.
	
Finally, following the numerical experiments above, and given the computational and time resources available, we conclude that with a size ``10", equal to $250$ travel requests of the randomly sampled sub-problems, the quality of the aggregate solution is, for the design nature of the problem, in line with its scope.

\subsection{Case Study of Manhattan}\label{Sec:Case}
In this section, we showcase our method using a data set with requests recorded on March 3, 2018. We randomly sample the data set 25 times with 250 requests each. We then solve Problem~\ref{prob:main} using the sampled data set to obtain a discretized distribution of the solution of the full problem. Solving each instance took a computational time of approximately \unit[3]{h}.
We highlight that economies of scales can significantly impact not only the solution, as discussed in Section~\ref{sec:sampleds}, but also the performance of an AMoD system~\cite{FielbaumTirachiniEtAl2021}. Two notorious examples are the Mohring and the Better-Matching effects. In bigger systems, larger number of vehicles and requests enable more efficient schedules and reduce delays. In this case in particular, the size of the sampled scenarios can impact the idling time, the rebalancing distance and the overall number of vehicles. 
\begin{table}[!t]
	\caption{Results}
	\label{tab:two}
	\begin{center}
		\begin{tabular}{|c||c||c||c|}
			\midrule
			Variable & Mean & St. Deviation & Unit \\
			\midrule
			$E^k_\mathrm{b}$ & $19.85$ & 2.11 & [kWh]\\
			\midrule
			$s^k$ & $3.44$ & 0.96 & [stops/vehicle]\\
			\midrule
			$B_\mathrm{r}^k $ & $24.53$ & 1.28 & [requests/vehicle]\\
			\midrule
			$C^k_\mathrm{avg}$ & $7.91$ & $7.29$& [kWh/stop]\\
			\midrule
			$r^k$ & $240.80 $ & $66.29$ & [€]\\
			\midrule
		\end{tabular}
	\end{center}
\end{table}
Table~\ref{tab:two} shows the optimal solution of Problem~\ref{prob:main}. Thereby, the battery size of the fleet, $E^k_\mathrm{b}$, is set to be equal between agents within the same batched problem. The number of recharging stops per vehicle is defined as 
\begin{equation}
	s^k = \sum_{i,j \in \cI^+, c \in \cC} S_{ijc}^k,
\end{equation}
while the number of requests served per vehicle $B^k_\mathrm{r}$, and the corresponding generated revenue $r^k$, are defined by
\begin{equation}
	B^k_\mathrm{r} = \sum_{i \in \pi(k)}b_\mathrm{r}^i,
\end{equation}
\begin{equation}
	r^k = \sum_{i \in \pi(k)}p_i b_\mathrm{r}^i,
\end{equation}
with $\pi(k)$ being the sequence of requests served by vehicle $k$. Last, 	$C^k_\mathrm{avg}$ is the average energy recharged per vehicle per stop.
The optimal battery size distribution lies near \unit[20]{kWh}, considerably smaller compared to commercial vehicles. This allows to  have a lower energy consumption thanks to lighter vehicles, thus reducing the electricity operational costs. The part of the solution related to the battery is consistent over different scenarios, as shown by the small standard deviations.
Regarding the charging scheduling of the fleet, each vehicle charges multiple times per day a small amount of energy, while only once, during night, a large amount, as indicated by the high standard deviation of $C^k_\mathrm{avg}$ in Table~\ref{tab:two}. This means that the vehicles have enough driving range to charge mainly whenever they are nearby a charging station, and when there are fewer requests.  
Interestingly, compared to~\cite{PaparellaHofmanEtAL2023}, where V2G was included, the battery is significantly downsized. In fact, because of the lack of incentives in charging and discharging during convenient time-widows, the battery is downsized so that the vehicles are as energy efficient and cheap as possible, but without sacrificing the potential number of travel requests that can be served.
In Section~\ref{sec:sensitivity} we will perform a sensitivity analysis on the battery size of the fleet. In particular, we solve Problem~\ref{prob:main} fixing the vehicle's design variables, hence removing the influence of the fixed costs per vehicle on the solution.

\subsection{Expected Lifetime of the Fleet}
\label{sec:lifetime}
A key aspect in this study is estimating the lifetime of the fleet, because it is the period over which the fixed costs are amortized. 
Specifically, the daily fixed costs are influenced by three elements: the cost of the single unit, the number of units and the expected lifetime. The cost of the single unit and the number of units depend on the battery capacity and operation of the fleet. The number of daily recharging cycles and the battery capacity influence the lifetime of the fleet itself.  
In particular, fleets with large battery capacity need less daily recharging cycles and are subject to lower degradation, reflecting in a longer lifetime. 

\begin{figure}[t]
	\centering
	\includegraphics[trim={0cm 0 1 10},clip,width=\columnwidth]{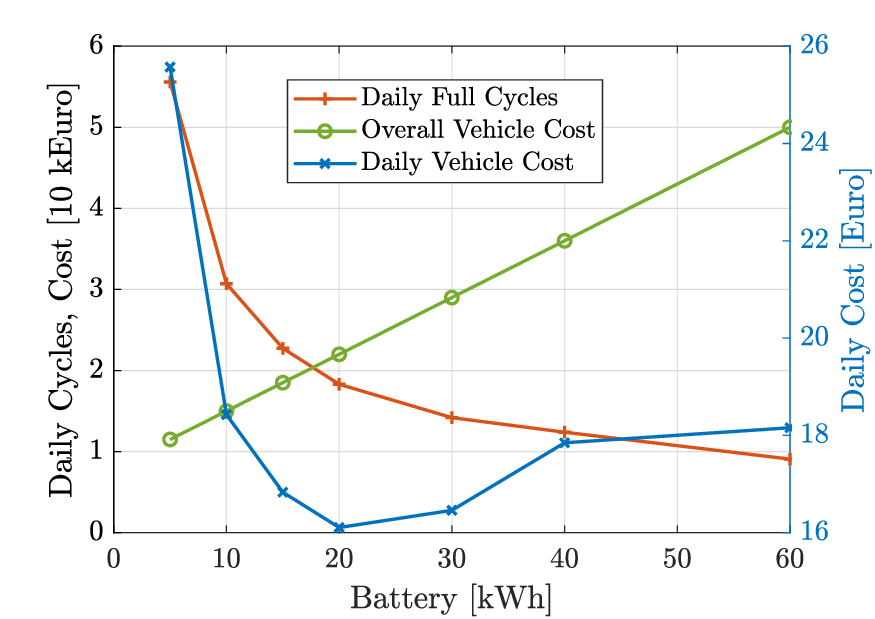}	
	\caption{Overall costs per unit in \unit[10]{kEuro} (red), daily cost per unit amortized over the lifetime (blue) and daily full battery cycles (yellow) as functions of the battery capacity of the unit. We recall that the minimum in the daily cost per unit is also due to the increasing energy consumption for vehicles with larger battery capacity.}
	\label{fig:Costs}
\end{figure}

We assume the lifetime bottleneck component of an electric vehicle to be the battery. We estimate the number of cycles before end of life of the battery to be 2500 full cycles. We assume the depth of discharge (DoD) does not significantly influence the battery's lifetime because the vehicles are charged for the most part by $20-40 \%$ per stop, i.e., the DoD is small.    
The amortization time in days for each vehicle is calculated by dividing the number of lifetime cycles by the daily ones.
Fig.~\ref{fig:Costs} shows the purchasing costs per unit and per unit amortized over its lifetime for different battery sizes. We highlight that battery degradation is not negligible, especially in vehicles with a very small battery capacity, that reflects in being more expensive on a daily basis, although being cheaper in absolute terms. However, for a battery size above \unit[10]{kWh}, the purchasing cost and degradation have comparable effects, resulting in an approximately constant daily costs. We also recall that we estimate the lifetime of the battery to be the bottleneck. This means that, for vehicles with very large battery capacity, the lifetime might be long enough to outlive the mechanical components, resulting in a possible overestimation of the lifetime of the vehicle. This could potentially reflect in a plateau of the lifetime, thus resulting in increasing amortized costs for vehicles with a large battery capacity, that are not captured in this model. Note that the lifetime is a function of the battery, which breaks the linearity of~\eqref{eq:obj}. However, we showed that above \unit[10]{kWh}, the purchasing costs are counterbalanced by the degradation of the battery, leading to an approximately constant daily cost per unit, independent of the battery. Therefore, to maintain a linear objective function, the terms $p_\mathrm{v}/\tau_\mathrm{v}$ and  $p_\mathrm{b}/\tau_\mathrm{v}$ are set to constant values. The approximation is valid only for battery size above \unit[10]{kWh}. However, in Section~\ref{sec:sensitivity} we will show that if the vehicles are equipped with a very small battery, a larger number of units will be required. Hence the solution does not lie in that part of the domain. Then, in Section~\ref{sec:sensitivity}, to gather a deeper insight into the effect of battery degradation on the solution, we perform a sensitivity analysis on the battery size, showing that it is not convenient to have a fleet with a very small battery size.

\subsection{Sensitivity Analysis on the Battery Size}\label{sec:sensitivity}
\begin{figure}[t]
	\centering
	\includegraphics[trim={0cm 10 30 1},clip,width=\columnwidth]{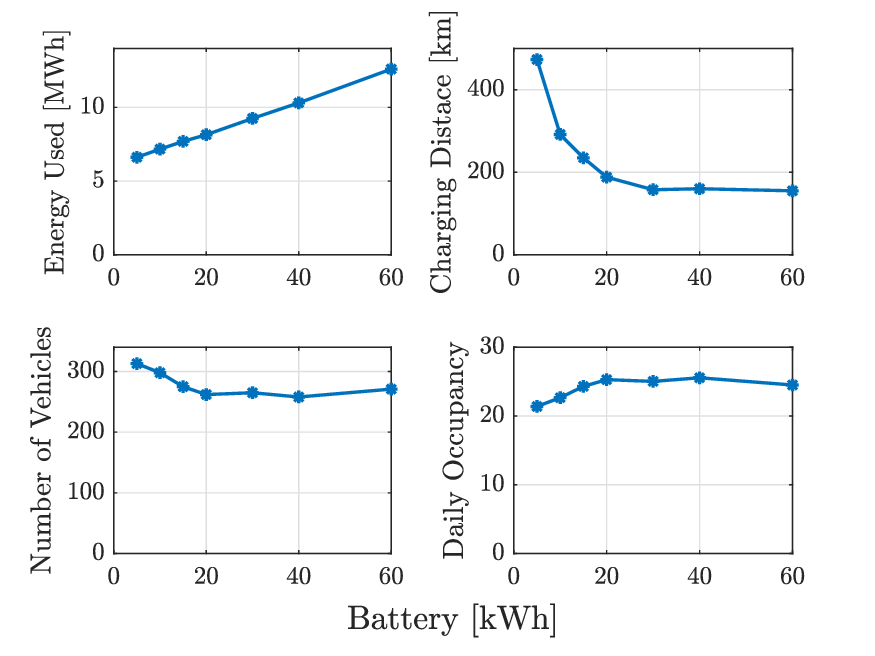}	
	\caption{The figure shows the energy usage (top-left), the additional driven to go to charging stations (top-right), the number of vehicles (bottom-right) and daily occupancy (bottom-left) as a function of the battery size of the vehicles of the fleet.} 
	\label{fig:quanti}
\end{figure}

\begin{figure}[t]
	\centering
	\includegraphics[trim={0cm 0 30 10},clip,width=\columnwidth]{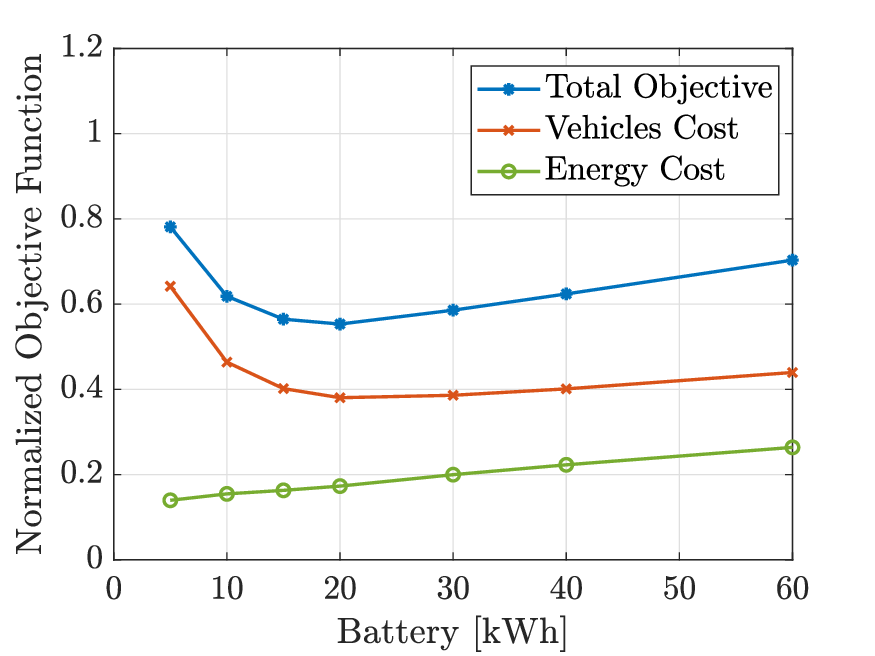}	
	\caption{The figure shows the sensitivity analysis of the objective function with respect to the battery size of the vehicles of the fleet. The number of served travel requests are fixed.} 
	\label{fig:obj}
\end{figure}

In this section we investigate how fleets with different battery capacities behave, in particular the trade-off between daily fixed costs and operational ones when fixing the number of daily requests served.
Fig.~\ref{fig:quanti} shows the relevant metrics of the system. For very small battery capacities, the additional distance driven to the respective charging stations is considerably higher, up to three times longer w.r.t. vehicles with larger batteries (top-right). The main reason is the absence of charging flexibility. In fact, the driving range is so short that the vehicles cannot wait to conveniently be near a charging station to recharge. Despite this, the order of magnitude of the additional distance driven is not comparable to the overall distance driven: This reflects in a lower energy consumption overall. From this we conclude that the operational costs are a monotone increasing function with respect to the battery size and weight of the single unit.
Conversely, the fixed amortized cost tends to be fairly constant for battery size above \unit[15]{kWh}, but it significantly increases in case of very small battery capacities, despite the lower purchasing cost per unit. The reasons are twofold. First, with a very small battery the flexibility of the charging schedule is absent. In this way, because of the continuous recharging process, in case of higher number of demands, the availability of vehicles cannot be temporarily increased, and more vehicles are required to serve the same number of requests. This leads to a larger, and thus more expensive, fleet. Second, a higher number of battery cycles leads to a shorter lifetime over which the fixed costs are amortized. Thus, both these reasons point to the fact that downsizing too much the battery is counter-productive. We also highlight that for very large batteries, the overall number of vehicles increases. The interpretation is that large vehicles consume more energy, leading to longer charging times, which, in turn, lead to lower availability of vehicles. To counter-act this phenomenon, the charging power can potentially be increased, especially in the case of vehicles with a large battery. However, for an easier comparison, we assume we have a fixed charging infrastructure.
Fig.~\ref{fig:obj} shows the trade-off between the fixed and the operational costs for a fixed number of travel requests served with an optimal value reached at approximately \unit[20]{kWh}. Around this value, the charging flexibility is high enough to not need additional vehicles to serve demand peaks, especially compared to lower battery capacities. In parallel, because of undersized batteries compared to the \unit[60]{kWh} ones, the single units are lighter and more energy efficient, reducing the overall energy consumption of the fleet.


%% file: Sections/Conclusions.tex
\section{Conclusions}\label{Sec:Concl}
In this paper, we proposed a framework for optimizing the design and operation of an Electric Autonomous Mobility-on-Demand (E-AMoD) fleet by considering the number of vehicles, battery size, and operation simultaneously. The framework combines vehicle assignment and charge scheduling with the design of the fleet.
To deal with the computational and combinatorial complexity of the resulting problem, we proposed a solution approach based on randomly sampling of the problem instance and, whenever needed, simplifying its structure via heuristic considerations.

We showcased our framework using a real-world case-study of Manhattan, NYC, showing that a fleet with battery size around \unit[20]{kWh} can i) result in a lower energy consumption without worsening the operation of the fleet compared to a fleet with a longer autonomy range, and ii) result in lower fixed costs compared to a fleet of vehicles with an undersized battery, which would have a shorter lifetime and require more vehicles due to the lower charging flexibility.

Going forward, our framework presents multiple avenues for expansion, such as the extension to ride-pooling, and to more sustainable intermodal settings that include public transit and active modes.

\section{Acknowledgments}\label{Sec:akn}
We thank Dr. I. New, Ir. M. Clemente and Ir. O. Borsboom for proofreading this paper, and Prof. A. Agazzi for the fruitful discussion and advice. 
This publication is part of the project NEON with project number 17628 of the research program Crossover which is (partly) financed by the Dutch Research Council (NWO). 